%% file: main.tex
\documentclass[aps,preprint,prx,longbibliography]{revtex4-1}
\pdfoutput=1
\usepackage[utf8]{inputenc}
\usepackage{amsmath,amssymb,mathrsfs,amsthm}
\usepackage{bbm}
\usepackage{chemarrow}
\usepackage{graphicx}
\usepackage{xcolor}
\usepackage{accents}
\usepackage{sidecap}
\input{macros}

\begin{document}

% title
\title{Neo-Gibbsian Statistical Energetics with Applications to Nonequilibrium Cells
}

\author{Bing Miao}
\affiliation{Center of Materials Science and Optoelectronics Engineering, College of Materials Science and Opto-Electronic Technology, University of Chinese Academy of Sciences, Beijing 100049, P.R.C.}

\author{Hong Qian}
\affiliation{Department of Applied Mathematics, University of Washington, Seattle, WA 98195, U.S.A.}

\author{Yong-Shi Wu}
\affiliation{Department of Physics \& Astronomy, University of Utah,\\ Salt Lake City, UT 84112, U.S.A.}

%%%%%%%%%%%%%%%%%% ABSTRACT %%%%%%%%%%%%%%%%%%

\begin{abstract}
Generalization through novel interpretations of the inner logic of the century-old Gibbs' statistical thermodynamics is presented: i)  Identifying $k_B\to 0$ as classical energetics, one directly derives a pair of thermodynamic variational formulae
\[
  F(T) = \min_{E\ge E_{min}}\Big\{E-TS(E) \Big\}
     \,\text{ and }\
  S(E) = \min_{T>0}\left\{\frac{E}{T}-\frac{F(T)}{T} \right\},
\]
that dictate all the more familiar $1/T=\rd S(E)/\rd E$, $E=\rd\{F(T)/T\}/\rd(1/T)$, and $S(E)=-\rd F(T)/\rd T$ in equilibrium, which is maintained by a duality symmetry with one-to-one relation between $T^{\text{eq}}(E)=\arg\min_T\{E/T-F(T)/T\}$ and $E^{\text{eq}}(T)=\arg\min_E\{E-TS(E)\}$.  ii)  In contradistinction, taking derivative of the statistical free energy w.r.t. $T$, a mesoscopic energetics with fluctuations emerges: This yields two information entropy functions which historically appeared 50 years postdate Gibbs' theory. iii) Combining the above pair of inequalities yields an irreversible thermodynamic potential $\psi(T,E) \equiv \{E-F(T)\}/T-S(E)\ge 0$ for nonequilibrium states.  The second law of thermodynamics as a universal principle reflects $\psi\ge 0$ due to a disagreement between $E$ and $T$ as a dual pair.  Our theory provides a new energetics of living cells which are nonequilibrium, complex entities under constant $T$, pressure $p$ and chemical potential $\mu$. $\psi$ provides a ``distance'' between statistical data from a large ensemble of cells and a set of intrinsic energetic parameters that encode the information within.
\end{abstract}

\maketitle

%\tableofcontents

%%%%%%%%%%%%%%%%%% INTRO %%%%%%%%%%%%%%%%%%

\section{Introduction}

Statistical thermodynamic energetics has served as the theoretical foundation for the biophysics of proteins with great success \cite{schellman_bpc_97}.  J. W. Gibbs was responsible for formulating the theory that included chemical molecules; it consists of both his method of modern thermodynamics \cite{guggenheim-book} and his statistical mechanics \cite{hill_ist}.  Currently one is witnessing a growing awareness of the importance of, and research excitement in, bioenergetics of cells \cite{physical_bioenergetics}; the latter as living organisms are nonequilibrium systems \cite{qian_jpc_06,WangJinRMP}. Still, there are hidden structures to be fully revealed in Gibbs' mathematical theory; what revealed below extend its scope to biological cells.

In protein research, J. Wyman developed a mathematical formulation for the statistical ligand binding to a protein with multiple sites in terms of a {\em binding polynomial} \cite{wyman} and further established the concept of binding potential that captures the thermodynamic linkage, or ``forces'', between sites and between binding and conformational transition \cite{wyman_gill}. This approach was behind the celebrated Monod-Wyman-Changeux model \cite{monod_wyman_changeux,hess_szabo}.  J. A. Schellman further discussed models in terms of statistical chemistry and thermodynamic approaches for the weak interaction between protein and cosolvent molecules \cite{schellman_75, Schellman_78}. He recognized the deep relation between his theory in biophysical chemistry and the generalized, isobaric grand canonical ensemble; but the latter had been known to be singular and non-existent in macroscopic thermodynamics since Gibbs' time, and statistical thermodynamics has been considered by some as an epistemological oxymoron \cite{schellman_bpc_97}.  Thanks to modern mathematics that refines Legendre transform into Legendre-Fenchel transform (LFT) \cite{rockafellar-book}, one now can legitimately express the LFT of an Eulerian homogeneous function as a function of all intensive conjugate variables with duality (see Eqs. \ref{TP-3} and \ref{SnLFT}).  This is precisely what needed for T. L. Hill's concept of subdivision potential in his thermodynamics of small systems \cite{hill_thermo_small} (see Eq. \ref{hgde} and Fig. \ref{fig02}).

In the current teaching of thermodynamics proper, the entropy concept originates either from analyzing the Carnot cycle {\em \'{a} la} R. Clausius, or from C. Carath\'{e}odory's inaccessibility principle \cite{lieb}, or else simply as a fundamental postulate stating its existence as a function of internal energy \cite{callen-book}.  One of the profound ideas of Gibbs on statistical thermodynamics was the mathematical equivalence, via Legendre transform, between the existence of an entropy as a concave function of internal energy, $S(E)$, and the existence of a free energy as a concave function of temperature, $F(T)$. Convexity defines thermodynamic phase and phase rule.  Further exploring this idea had led to the ``crucial step'' for T. L. Hill formulating his thermodynamics of small systems without randomness in 1960s \cite{hill_nanolett,lu-qian-22}.

The Legendre transform, or in modern terminology Legendre-Fenchel (LF) dual relation \cite{rockafellar-book}, was prominent in classical thermodynamics, but it is conspicuously absent in the current teaching of statistical mechanics; the LF duality {\em symmetry} is completely missing. In the present work, we shall elevate LF duality symmetry to the defining feature of thermodynamic equilibrium; the duality symmetry breaking implies nonequilibrium \cite{MQW-I}.

This paper is structured as follows: In Sec. \ref{sec:2} we have found a clear conception for Helmholtz free energy $F(T)=E-TS(E)$ as an equilibrium relationship in which $E$ is not arbitrary when $T$ is given and {\em vice versa}, and Helmholtz free energy $E-TS(E)$ as a nonequilibrium thermodynamic potential function of a given $T$ and varying $E$. These two interpretations are related via a variational relation $F(T)=\min_{E}\big\{E-TS(E)\big\}$, which finds the classical equilibrium $E$ corresponding to the given $T$.  We obtain this result through carefully crafting the thermodynamic limit $(k_B\to 0)$ using some new but elementary mathematics that is fully accessible to both experimental and theoretical biophysicists at the level of \cite{BenNaimBook,DillBook,BarrickBook}, with only new mathematical expressions as in Eqs. (\ref{FT}) and (\ref{SE}). In Sec. \ref{sec:3}, by taking $\rd/\rd T$ before thermodynamic limit $k_B\to 0$, we discover that macroscopic and mesoscopic thermodynamics are fundamentally two distinct theories {\em \'{a} la} P. W. Anderson \cite{anderson72}.  Sec. \ref{sec:4} further shows that the concept of irreversible thermodynamic potential naturally arises {\em \`{a} la} the Brussels school of irreversible thermodynamics; this is consistent with that the second law not being founded on ``the arrow of time'' forcefully articulated in \cite{lieb}.

These above results, when combined with the recent, complementary studies  \cite{qian_jctc,qian_entropy_24,cqy_cpb}, clearly show that energetics as Gibbsian statistical thermodynamic concepts can be applied to statistical data from nonequilibrium systems like living cells.  We give an outline of this provocative idea in Sec. \ref{sec:5}, more comprehensive exposition is forthcoming.

Our new identification of $k_B\to 0$ as classical thermodynamic limit immediately allows the conception of classical thermodynamics of macroscopic extensive systems as well as small systems \cite{hill_thermo_small}.  $k_B\to 0$ as classical thermodynamics is to thermal fluctuations what $\hbar\to 0$ as classical mechanics to quantum fluctuations \cite{planck_1900_2}. One notices that the Boltzmann constant $k_B$ serves as the unit for entropy, the central object representing thermal fluctuations in statistical physics; it is in parallel to the Planck constant $\hbar$ as the unit for action that is central to studying quantum fluctuations.  Compared to the traditional thermodynamic limit with system's size $N,V\to \infty$ while holding $N/V$ constant, the vanishing fluctuation applies to both large and small systems.\footnote{Using path integrals representing time-dependent fluctuations, either thermal or quantum, the operations of $k_B\to 0$ and $\hbar\to 0$ single out, respectively, the path with maximum entropy and minimum action as dominant contribution to the propagator, resulting in their corresponding variational principles. This is in fact the spirit of the WKB method in semiclassical theories \cite{MQW-II}. }

Just as protein science is founded partly on Gibbs' theory \cite{edsall-book,DillBook,BarrickBook}, the science of living cells cannot avoid statistical physics.  The opening paragraph of Landau and Lifshitz's \cite{landau_lifshitz},  with only minor amendments as marked, articulates very well for our work overall:
\begin{quote}
{\em Statistical physics}, often called for brevity simply {\em statistics}, consists in the study of the special laws which govern the behavior and properties of macroscopic bodies (that is, bodies formed of a very large number of individual particles [or events], such as atoms and molecules, [or data]).  To a considerable extent the general character of these laws does not depend on the mechanics (classical or quantum) which describes the motion of the individual particles in a body, [or the mechanisms underlying the data], but their substantiation demands a different argument in the [three] cases.
\end{quote}

\noindent
For classical mechanical systems, the laws as formulated by Gibbs are encapsulated in Eq. (\ref{BoEn}) below, which begins with a given mechanical energy $U(\vx)$ and a temperature $T$.
See \cite{qian_jctc,qian_entropy_24} for the laws of cellular data {\em ad infinitum} that starts from the theory of probability.

\section{Classical thermodynamics via variational principle as $\boldsymbol{k_B\to 0}$}
\label{sec:2}

Let us use $\mathcal{S}\subset\mathbb{R}^N$ to denote the state space of a ``microscopic, complete'' description of a system with an energy function $U(\vx)$, $\vx\in\mathcal{S}$.  We deliberately keep the nature of $\vx$ vague: the state and energy function can be Newtonian mechanical, or chemical, or even something more general and elusive.  Such approach is not foreign to a biophysical chemist who has been exposed to the theory of helix-coil transition for understanding the secondary structure within a polypeptide \cite{Schellman_1958}.

\begin{figure}[t]
\includegraphics[scale=.3]{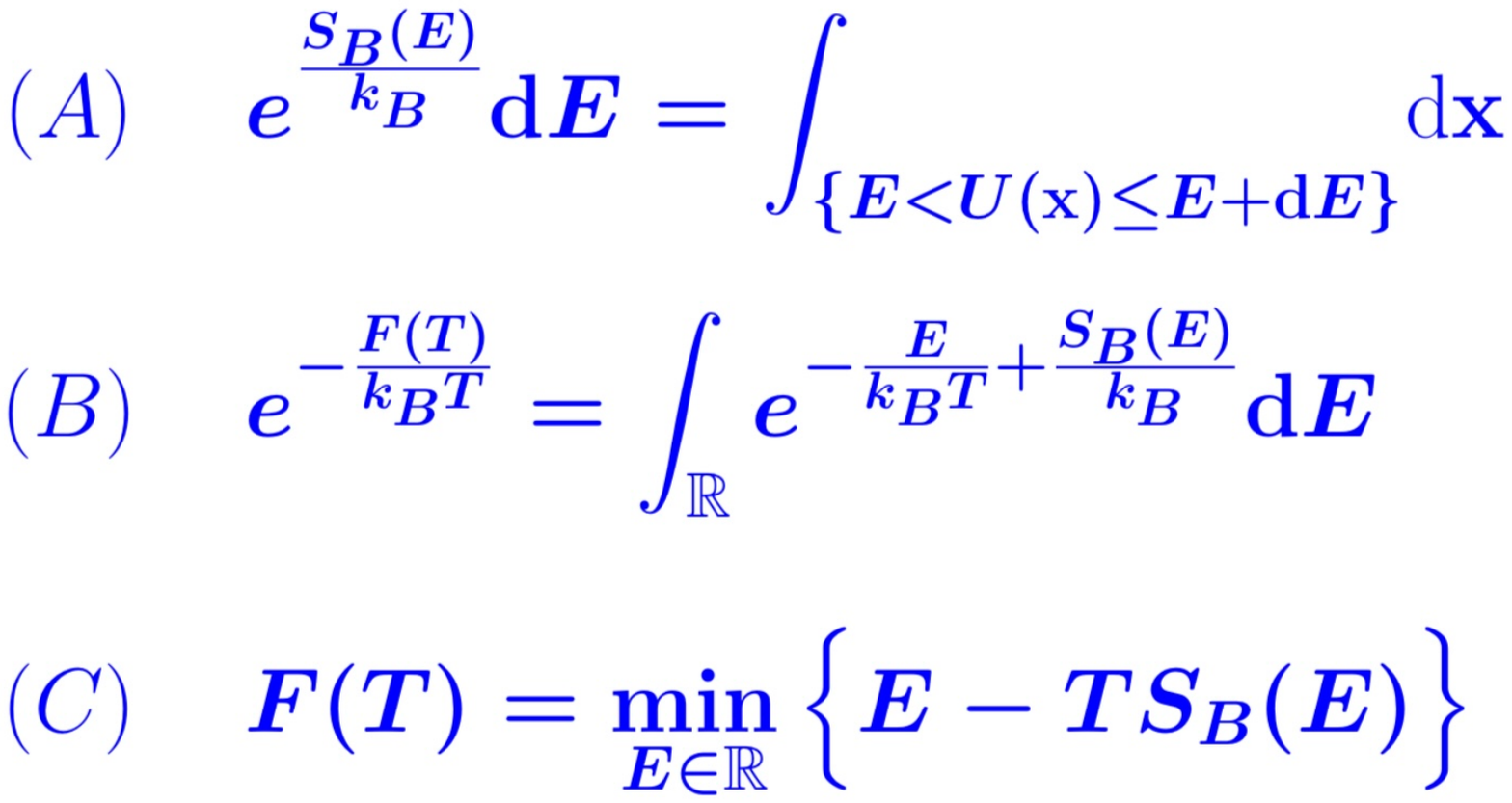}
\caption{Three essential equations that serve as the law of statistical energetics replacing the {\em fundamental thermodynamic relation}; the first two are found in any statistical mechanics textbook in which the third is likely to be missing.  Eq. A is due to Boltzmann; it relates a given energy function $U(\vx)$ of a state space point $\vx\in\mathcal{S}$ to the entropy function $S_B(E)$ as the logarithm of the volume, inside state space $\mathcal{S}$, contained by all the $\vx$ whose $U(\vx)\in (E,E+\rd E]$; temperature $T$ is not involved.  Eq. B is due to Gibbs; it relates $S_B(E)$ to free energy $F(T)$.  The integral on the right is known as a partition function, which can be expressed equivalently as the integration of $e^{-U(\vx)/k_BT}$ over an appropriate $\mathcal{S}$ (Eq. \ref{BoEn}b).  In the thermodynamics limit with $F$, $E$, and $S_B$ all grow linearly with the size of a system, most textbooks deduce from Eq. B that $F(E)=E-TS_B(E)$ in thermodynamic {\em equilibrium}.  The present work suggests that as $k_B\to 0$, Eq. C is the rigorous mathematical consequence of Eq. B, together with $1/T^{\text{eq}}=\rd S_B(E)/\rd E$.  Eq. C implies $F(T)\le E-TS_B(E)$ for any arbitrary value of $E$ under a given $T$, and in equilibrium $E^{\text{eq}}=\rd\{F(T)/T\}/\rd(1/T)$.
}
\label{fig01}
\end{figure}

As summarized in Fig. \ref{fig01}, we start from Boltzmann's entropy function
\begin{subequations}
\label{BoEn}
\begin{equation}
     S_B(E) = k_B\log\left(\frac{1}{\rd E} \int_{
     \{E<U(\vx)\le E+\rd E\}\bigcap\mathcal{S} }
       \rd\vx \right),
\end{equation}
and Helmholtz free energy function according to Gibbs
\begin{eqnarray}
    F(T) &=& -k_BT\log\int_{\mathcal{S}}
       e^{-\frac{U(\vx)}{k_BT}} \, \rd\vx
\\
     &=&  -k_BT\log\int_{E_{min}}^{+\infty}  e^{-\frac{E}{k_BT}}\,
     \int_{\{E<U(\vx)\le E+\rd E\}\bigcap\mathcal{S}} \rd\vx
\nonumber\\
    &=& -k_BT\log\int_{\mathbb{R}}  e^{-\frac{E/T-S_B(E)}{k_B}} \, \rd E.
\end{eqnarray}
The set $\{E<U( \vx)\le E+\rd E\}$ denotes all the $\vx\in\mathbb{R}^N$ that has energy $U(\vx)\in (E,E+\rd E]$.   One can find this mathematical writing, which is different from Boltzmann's counting of discrete states, in \cite{khinchin-book}.  We assume the energy function is bounded from below $U(\vx)\ge E_{min}\equiv \min_{\vx\in\mathcal{S}} U(\vx)$.
In (\ref{BoEn}c), $S_B(E)=-\infty$, $e^{S_B(E)/k_B}=0$ for all $E<E_{min}$, and $\rd S_B(E)/\rd E=+\infty$ at $E=E_{min}$.

\end{subequations}

The notion of an extensive variable/quantity is fundamental in classical thermodynamics \cite{callen-book}.  However from a more careful mathematical standpoint, such quantities do not exist in the thermodynamic limit. We shall instead use $k_B\to 0$ to represent {\em classical macroscopic thermodynamic limit} of energetic quantities that include entropy and free energy. In Boltzmann's work $k_B$ represents the conversion between the kinetic energy and the entropy based on particle counting, thus $k_B\to 0$ is consistent with the number of particles tends to infinity.  We shall clearly see from the analysis below that this choice for connecting statistical mechanics to classical thermodynamics is in the same spirit as the using of $\hbar\to 0$ for establishing the logical relation between quantum and classical mechanics.

Furthermore, we shall show that the very existence of thermodynamic potential functions and related variational principles are merely the mathematical consequence of the limiting process.  This new formulation perfectly echos a recent understanding of entropy through the modern theory of probability and its large deviations \cite{qian_jctc,qian_entropy_24}.

From Eq. (\ref{BoEn}c) one immediately has
in the limit of $k_B\to 0$:
\begin{equation}
  F(T) = \min_{E\in\mathbb{R}}\Big\{ E-TS_B(E)\Big\},
\label{FT}
\end{equation}
in which $T>0$, $E\in\mathbb{R}$ and $E\ge E_{min}$ are equivalent since $S_B(E)=-\infty$ for $E<E_{min}$. The LF dual relation leads to a different $S(E)$:
\begin{equation}
\label{SE}
    S(E) = \min_{T>0}\left\{\frac{E}{T} -\frac{F(T)}{T} \right\}.
\end{equation}
Both $F(T)$ and $S(E)$ are concave functions by their very definitions in Eqs. (\ref{FT}) and (\ref{SE}).  If $S_B(E)$ in (\ref{BoEn}a) is itself a concave function of $E$, then there is an LF duality symmetry $S_B(E)=S(E)$ \cite{lu-qian-22}.  In general, however, $S_B(E)\le S(E)$ which is the non-smooth concave hull of $S_B(E)$.  A non-concave $S_B(E)$ with multiple concave domains leads to phase transition scenario \cite{qian_cpl_17}. The overall equilibrium is represented by the larger $S(E)$.

Eqs. (\ref{FT}) and (\ref{SE}) are a pair of thermodynamic variational formulae. On the rhs of (\ref{FT}), $E-TS_{B}(E)$ should be identified, macroscopically, as a nonequilibrium free energy function under a fixed $T$ for all possible $E$'s.  The equilibrium $E^{\text{eq}}$ corresponding to the $T$ is the minimizer of the nonequilibrium free energy: $E^{\text{eq}}(T)=\arg\min_E\{E-TS_B(E)\}$. At a finite $k_BT$, however, (\ref{BoEn}c) shows that the same free energy function governs the distribution of fluctuating $E$.  This is a necessary but not sufficient condition for a mesoscopic, statistical equilibrium over the entire state space $\mathcal{S}$.  Similarly, the rhs of (\ref{SE}) is another nonequilibrium potential under a given $E$ for all possible $T$'s: the equilibrium $T^{\text{eq}}(E)=\arg\min_T\{E/T-F(T)/T\}$.

For concave $S_B(E)$, function $E^{\text{eq}}(T)$ is the inverse function of $T^{\text{eq}}(E)$.  Both nonequilibrium entropy and free energy functions are examples of {\em thermodynamic potentials}, a concept that has been widely suggested but was never made clear in mathematical terms \cite{landau_lifshitz}.
The duality has led to the notion of temperature fluctuation \cite{mandelbrot_89}.  Note that equilibrium Helmholtz free energy $F(T)$ and nonequilibrium free energy $E-TS_B(E)\ge F(T)$ are two different thermodynamic concepts.  Similarly, equilibrium $S(E)$ and nonequilibrium $E/T-F(T)/T\ge S(E)$ are two distinct concepts.  Following this logic, the Gibbs potential widely employed in biophysical chemistry is $E+pV-TS_B(E,V,n)\ge G(T,p,n)$ for fixed $T,p$ and varying $E,V$.  See Appendix \ref{app_A} for more discussion.

Computationally the minimum of Eq. (\ref{FT}) as well as the minimization in Eq. (\ref{SE}) can be found using calculus.  They imply
\begin{equation}
\label{LFsys}
    \frac{1}{T} = \frac{\rd S(E)}{\rd E },\,
    E=\frac{\rd \{F(T)/T\} }{ \rd(1/T)},
        \, \text{ and }  \,
    S(E)=-\frac{\rd F(T)}{\rd T}.
\end{equation}
These are celebrated macroscopic, equilibrium thermodynamic relations.  Furthermore,
\begin{equation}
    \frac{\rd^2 S(E)}{\rd E^2} = \frac{\rd(1/T)}{\rd E} \, \text{ and } \,
      \frac{\rd^2\{F(T)/T\}}{\rd(1/T)^2}
      =\frac{\rd E}{\rd(1/T)}=-T^2C< 0,
\end{equation}
representing convexity and variational stability.  It is now clear that all these relations are consequences of the variational relations in (\ref{FT}) and (\ref{SE}), which in turn are derived from Eq. (\ref{BoEn}), according to the mathematical procedure
\begin{equation}
     k_B \to 0 \,\text{ first, followed by } \
     \frac{\rd}{\rd T} \,\text{ and }\,
     \frac{\rd}{\rd E}.
\label{macro_limit}
\end{equation}
This procedure holds macroscopic entropy $S_B(E)$ constant as $k_B\to 0$; as shown in Appendix \ref{app_A}, it is formally equivalent to the macroscopic limit with infinite degrees of freedom.

\section{Mesoscopic thermodynamic entropy $\boldsymbol{-\rd F/\rd T}$}
\label{sec:3}

The last relation in (\ref{LFsys}) derives classical thermodynamic entropy as the derivative of free energy w.r.t. $T$.  Exchanging the order in Eq. (\ref{macro_limit}) might yield a different result.  It is well-known in condensed matter physics that such a difference defines {\em emergent phenomenon} \cite{qian_cpl_17,Chibbaro-book}.  Carrying out derivative w.r.t. $T$ for the $F(T)$ in (\ref{BoEn}b) one has
\begin{subequations}
\label{dFdT}
\begin{eqnarray}
    -\frac{\rd F(T)}{\rd T}
    = -k_B\int_{\mathcal{S}}
      P(\vx)\log P(\vx) \, \rd\vx
\nonumber\\
  \text{ where }  \,
    P(\vx) = \frac{ e^{-U(\vx)/k_BT} }{\displaystyle
       \int_{\mathcal{S}} e^{-U(\vx)/k_BT}\rd\vx  },
\end{eqnarray}
which can be further expressed in two more meaningful forms:
\begin{eqnarray}
 -\frac{\rd F(T)}{\rd T}
     &=& \underbrace{ -k_B\int_{\mathcal{S}}
      P(\vx)\log\Big(P(\vx)\rd\vx\Big) \, \rd\vx
       }_{\ge\, 0}+ k_B\log\big(\rd\vx\big)
\nonumber\\[-14pt]
\\
    &=& \underbrace{ -k_B\int_{\mathcal{S}}
      P(\vx)\log\left(\frac{ P(\vx) }{\|\mathcal{S}\|^{-1} } \right) \, \rd\vx
       }_{\le \, 0}+ k_B\log\|\mathcal{S}\|,
\nonumber\\[-14pt]
\\
    && \text{in which }\,
     \int_{\mathcal{S}} P(\vx)\rd\vx = \int_{\mathcal{S}} \|\mathcal{S}\|^{-1}\,\rd\vx
    = 1.
\nonumber
\end{eqnarray}
\end{subequations}
Without taking the $k_B\to 0$ limit, $-\rd F/\rd T$ in (\ref{dFdT}) is naturally identified as {\em mesoscopic thermodynamic entropy}.  More importantly, being the same mathematical quantities in two different expressions, Eqs. (\ref{dFdT}b) and (\ref{dFdT}c) clearly tell us that entropy is only meaningful w.r.t. a choice of a reference:  It can be either positive w.r.t state volume element $\log(\rd\vx)$ or negative w.r.t. total state space $\log\|\mathcal{S}\|$. We observe that if one discretizes the $\mathcal{S}$ with resolution $\rd\vx$ in (\ref{dFdT}b), then $P(\vx)\rd\vx \to P_i$, $i=1,2,\cdots$, which is the celebrated Shannon entropy in units of $k_B$.  Thus Eqs. (\ref{dFdT}b) and (\ref{dFdT}c) are the two most important expressions in information theory \cite{cover-book}, Shannon's entropy and relative entropy, also known as Kullback-Leibler divergence w.r.t. uniform distribution \cite{shore-johnson}.  Therefore it is reasonable to conclude that from a strict scientific logic, information theory had been completely anticipated by Gibbs' theory as an alternative treatment in Sec. \ref{sec:2}; a completely legitimate derivative of statistical chemistry.

Noticing the $k_B$ in Eqs. (\ref{dFdT}b,c), if one takes $k_B\to 0$ following $\rd F/\rd T$, the $P(\vx)$ in Eq. (\ref{dFdT}a)
can be represented by a Gaussian distribution centered around the global minimum of $U(\vx)$ with variance $\sigma^2\propto k_BT$:
\[
  P(\vx) =  \frac{1}{\sqrt{2\pi\sigma^2}}
                e^{-\frac{(\vx-\vx^*)^2}{2\sigma^2}}, \ \
                \vx^* = \arg\min_{\vx\in\mathcal{S}} U(\vx).
\]
Therefore,
\begin{equation}
    k_B\int_{\mathbb{R}} P(x)\log P(x) \rd x
    = k_B\left(-\frac{1}{2}-\log\sqrt{2\pi\sigma^2 }\right) \to 0
\end{equation}
as $k_B\to 0$ following $\rd/\rd T$.  Without holding a finite entropy as $k_B\to 0$, the
mesoscopic approach yields zero entropy as $k_B\to 0$; and the discrepancy can be interpreted as in the reference $k_B\log(\rd\vx)$ or $k_B\log\|\mathcal{S}\|$.  The disagreement with the ``correct'' macroscopic thermodynamics as given in Sec. \ref{sec:2}, however, is scientifically more significant because it is a new theory for mesoscopic thermodynamic energetics with fluctuations.  It had been a prophecy for the rise of information theory that came many years later \cite{cover-book} and even for the current stochastic nonequilibrium thermodynamics proper \cite{peliti-book,shiraishi-book}, as shown next.

\section{Irreversible thermodynamic potential}
\label{sec:4}

In classical thermodynamic equilibrium there is an LF duality symmetry relation between $S(E)$ and $F(T)$.  However, as pointed out above the right-hand-sides of (\ref{FT}) and (\ref{SE}) inside the $\{\cdots\}$ are actually macroscopic irreversible thermodynamic potential functions, for given $T$ and given $E$, respectively.  Combining the two expressions, one obtains a single inequality, known as the Fenchel-Young inequality \cite{rockafellar-book}
\begin{equation}
     \frac{E}{T}-S(E)-\frac{F(T)}{T} \ge 0,
\end{equation}
with the equal sign holding true if and only if the equilibrium relation in Eq. (\ref{LFsys}) being valid: There is a one-to-one relation between $T$ and $E$ in equilibrium; either one of them can be used to specify a macroscopic equilibrium state in a laboratory: isothermal vs. constant energy.  A nonequilibrium state is implied when its $T$ and $E$ do not satisfy Eq. (\ref{LFsys}).  Therefore it is very natural to introduce the concept of {\em irreversible thermodynamic potential} w.r.t. equilibrium \cite{MQW-I}:
\begin{subequations}
\label{ep}
\begin{eqnarray}
   \psi(T,E) &=& \frac{1}{T} \underbrace{  \Big\{ E-T S(E)\Big\} }_{\text{non-eq. free energy}} -\min_{E\in\mathbb{R}}
            \left\{ \frac{E}{T} - S(E)\right\}
\\
   &=& \max_{T\ge 0}
             \left\{ \frac{F(T)}{T}
             -\frac{E}{T}\right\} - \underbrace{ \frac{F(T)-E}{T}  }_{\text{non-eq. entropy}}
             \ \ge 0,
\end{eqnarray}
\end{subequations}
$\psi=0$ represents an equilibrium state, and $\psi>0$ signifies a nonequilibrium state.  In fact the inequality $\psi>0$ is a form of the Second Law of Thermodynamics, without a reference to the ``arrow of time'' \cite{qian_jctc,lieb} nor dynamics \cite{mackey_rmp}.  It is a purely statistical law that emerges in the thermodynamic limit.  Eqs. (\ref{ep}a) and (\ref{ep}b) are traditionally discussed, separately, as canonical system with fixed $T$ and micro-canonical system with fixed $E$. That nonequilibrium free energy and nonequilibrium entropy can be combined into a unified formulation was an insight of the Brussels school of irreversible thermodynamics through entropy production \cite{prigogine-book}.

For mesoscopic systems with notable fluctuations,  $\psi(T,E)=0$, which represents a macroscopic equilibrium between temperature and energy, is no longer meaningful. The detailed description in terms of statistical distribution at $T$ matters.  When substituting statistical expression for $F(T)$ in (\ref{BoEn}b), together with mesoscopic entropy and energy as
\[
    S = -k_B\int_{\mathcal{S}} P(\vx)\log P(\vx) \rd\vx, \ \
    E = \int_{\mathcal{S}} U(\vx)P(\vx)\rd\vx,
\]
into Eq. (\ref{ep}), one immediately obtains a mesoscopic correspondence \cite{qian_jctc}:
\begin{subequations}
\label{meso_neq_p}
\begin{eqnarray}
    \psi^{(\text{meso})} &=& \frac{1}{T} \int_{\mathcal{S}} U(\vx)P(\vx)\rd\vx + k_B\int_{\mathcal{S}}
      P(\vx)\log P(\vx) \rd\vx
    +k_B\log\int_{\mathcal{S}} e^{-\frac{U(\vx)}{k_BT}}\rd\vx
\\
    &=& k_B\left\{  \int_{\mathcal{S}}
      P(\vx)\log P(\vx) \rd\vx - \int_{\mathcal{S}} P(\vx) \log\frac{ e^{-\frac{U(\vx)}{k_BT}} }{\displaystyle
      \int_{\mathcal{S}} e^{-\frac{U(\vx)}{k_BT}}\rd\vx}\rd\vx
      \right\}
\nonumber\\
    &=& k_B\int_{\mathcal{S}}
      P(\vx)\log\left(\frac{ P(\vx) }{P^{\text{eq}}(\vx)}\right)\rd\vx, \text{ where }
       P^{\text{eq}}(\vx) = \frac{ e^{-\frac{U(\vx)}{k_BT}} }{\displaystyle
      \int_{\mathcal{S}} e^{-\frac{U(\vx)}{k_BT}}\rd\vx}.
\end{eqnarray}
\end{subequations}
The last line indicates it is a Kullback-Leibler divergence w.r.t. the mesoscopic, statistical equilibrium distribution $P^{\text{eq}}(\vx)$. Therefore it is non-negative and goes to zero when $P(\vx)$ equals to $P^{\text{eq}}(\vx)$.  A mesoscopic nonequilibrium system means $P(\vx)\neq P^{\text{eq}}(\vx)$.  The first two terms in (\ref{meso_neq_p}a) are mean internal energy and entropy; together they form the nonequilibrium, irreversible free energy for varying $P(\vx)$.  The last term is the equilibrium free energy as its minimum among all possible $P(\vx)$; it is the LF transform of the mesoscopic entropy with $U(\vx)$ and $P(\vx)$ being conjugate pair:
\begin{equation}
    \min_{P} \left\{  \int_{\mathcal{S}} U(\vx)P(\vx)\rd\vx + k_BT\int_{\mathcal{S}}
      P(\vx)\log P(\vx) \rd\vx
    \right\} = -k_BT\log\int_{\mathcal{S}} e^{-\frac{U(\vx)}{k_BT}}\rd\vx.
\end{equation}
Gibbs' statistical mechanics is merely a mesoscopic thermodynamic theory with the state space consisting of all possible distributions on $\mathcal{S}$. Thermodynamics is more fundamental than statistical mechanics \cite{qian_jctc,qian_entropy_24}.

\begin{figure}[t]
\includegraphics[scale=.55]{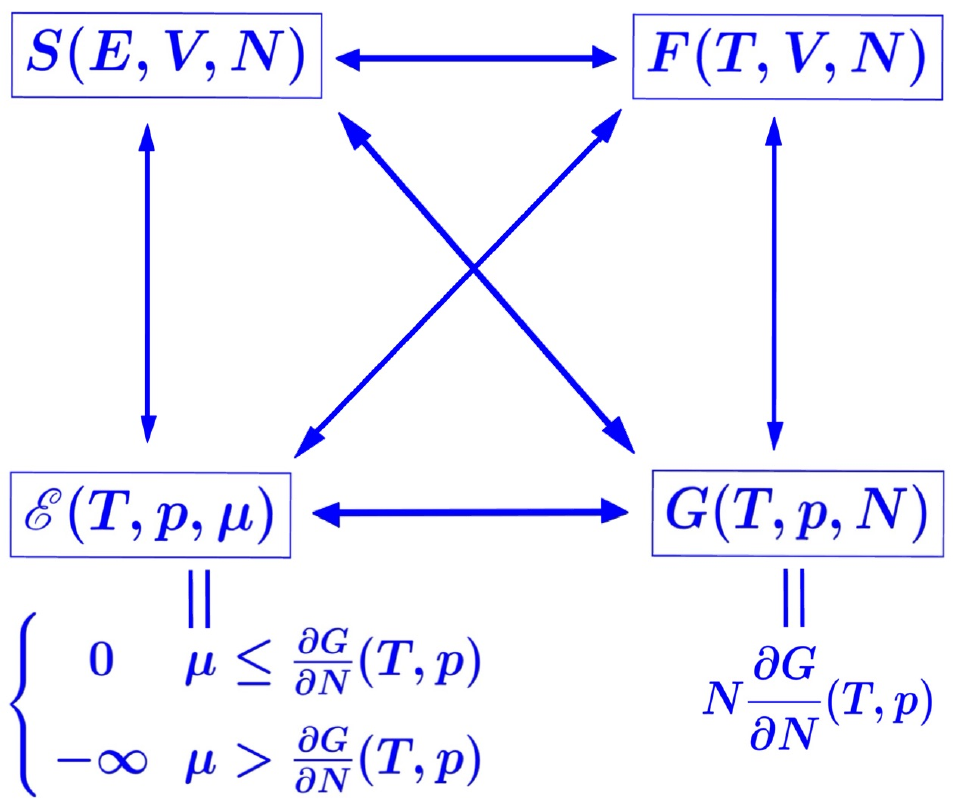}
\caption{The structure of statistical energetics. Various thermodynamic potential functions are shown.  Classical mechanics yields entropy $S(E,V,N)$ {\em \`{a} la} Boltzmann; the most appropriate for a biological system under room temperature $T$, $1$ atm $p$, and given chemical potential $\mu$ is Hill's subdivision potential $\mathscr{E}(T,p,\mu)$, which is known as isobaric grand canonical system \cite{schellman_75}.  If $F(T,V,N)$ is a concave function, all  other potential functions can be derived bi-directionally: $S(E,V,N)=\min_{T}\{[E-F(T,V,N)]/T\}$ and $F(T,V,N)=\min_E\{E-TS(E,V,N)\}$.  Similarly, $\mathscr{E}(T,p,\mu)=\min_{E,V,N}\{ E+pV-\mu N-TS(E,V,N)\}$ $=\min_{V,N}\{pV-\mu N+F(T,V,N)\}$ $=\min_{N}\{-\mu N+G(T,p,N)\}$, and conversely $S(E,V,N)=\min_{T,p,\mu}\{ (E+pV-\mu N-\mathscr{E}(T,p,\mu))/T\}$ $=\min_{T,p}\{(E+pV-G(T,p,N))/T\}$.  We do not list exhaustively all the other dual relations to avoid cluttering.  Classical thermodynamics can be obeyed for both macroscopic and small systems \cite{hill_thermo_small}, and it needs not to be tied to mechanics. Only for a macroscopic system, Gibbs potential $G(T,p,N)=N(\partial G/\partial N)(T,p)$ is a linear function of $N$, and correspondingly $\mathscr{E}(T,p,\mu)$ is singular as shown in the figure but it actually exists mathematically as a legitimate concave function, and a duality holds so that $G(T,p,N)=\max_{\mu}\{\mu N+\mathscr{E}(T,p,\mu) \}$.
}
\label{fig02}
\end{figure}

\section{Single-cell high dimensional data}
\label{sec:5}

Fig. \ref{fig02} succinctly captures the discussion above.  Our neo-Gibbsian theory of statistical energetics presented echos a completely parallel logic based on the large deviations in the mathematical theory of probability \cite{qian_jctc,qian_entropy_24}, especially Cram\'{e}r's theorem and Sanov's theorem \cite{dembo-book}, applied to empirical data. With this understanding, it is applicable to any set of measurements idealized as recurrent and infinitely large |  data {\em ad infinitum}.  In this case, internal energy function $U(\vx)$ is no longer considered as mechanical, nor fundamental.  Rather they are parameters, e.g. statistical weights, in a statistical model, under a given probability $P(\vx)$, that accounts for observed statistical counting frequencies.  Expressed for a set of discrete states $k\in\mathcal{S}\equiv\{1,\cdots,K\}$ under probability $\{p_k\}$ and a large set of $N$ samples, among them $n_k$ samples are in state $k$:
\begin{equation}
     \frac{n_k}{N} = p_k e^{-G_k}.
\end{equation}
In energy units of $k_BT$, this is precisely the practice in biophysical chemistry: Measuring biochemical concentrations and converting the results into energetics:
\begin{equation}
     G_i-G_j = \log\left(\frac{p_i}{p_j}\right) -\log\left(\frac{n_i}{n_j}\right) = \Delta G_{ji}^o-\log\left(\frac{n_i}{n_j}\right),
\end{equation}
which is further split into entropic and enthalpic components via Van't Hoff relation.

Whether they are a population of classified biological macromolecules like proteins or living cells, this approach can be completely adopted for analyzing data from single cells \cite{qian-cheng-qb}: Under the assumption of independent and identically distributed (i.i.d.) individual cells, there is an entropy function of counting $n_i$, $i\in\mathcal{S}$, that is Eulerian degree $1$ homogeneous:
\begin{equation}
\label{Sn}
    S(\vn) = \sum_{i=1}^K n_i\log
    \left(\frac{n_i}{(n_1+\cdots+n_K)p_i}
    \right),
\end{equation}
whose full LF transform yields the ``singular'' function but it exists:
\begin{equation}
\label{SnLFT}
     \max_{\vn}\left\{
    \sum_{i=1}^K n_iu_i -S(\vn)\right\}
    = \left\{ \begin{array}{ccc}
        0 &&  \hat{F}(\vu) \le 0  \\
        +\infty && \hat{F}(\vu) > 0
    \end{array} \right.,
\end{equation}
where $N\hat{F}(\vu)=F(\vu)$, the free energy with fixed $N$
\begin{eqnarray}
    &\displaystyle
    F(\vu) = \max_{\vn}\left\{
    \sum_{i=1}^K n_iu_i -S(\vn)\left|
    \, \sum_{i=1}^K n_i = N
    \right.\right\}
    = N\log\sum_{i=1}^K p_ie^{u_i}, \hspace{.2in}
\label{Fu}\\
    &\displaystyle
    \arg\max_{\vn}\left\{
    \sum_{i=1}^K n_iu_i -S(\vn)\left| \, \sum_{i=1}^K n_i = N
    \right. \right\}
    = \frac{Np_ie^{u_i}}{\displaystyle
    \sum_{i=1}^K p_ie^{u_i}}, \
    i\in\mathcal{S}. \hspace{.3in}
\end{eqnarray}
One only has to identify $u_i$ as negative energy in $k_BT$ units to see the parallel to the biophysics of dilute solution of proteins.  It should also be noted that $S(\vn)$ and $F(\vu)$ in (\ref{Sn}) and (\ref{Fu}) are discrete-state counterparts of the entropy in (\ref{dFdT}a) and free energy in (\ref{BoEn}b), in $k_BT$ units.

With a discrete cellular state space $\mathcal{S}$ given, $L$ different measurements from a single cell are considered functions on the state space $\mathcal{S}$: Each state $k\in\mathcal{S}$ yields a unique set of
measurements $x^{(k)}_{\ell}$, $1\le\ell\le L$.  We call $x_{\ell}^{(k)}$ the $\ell^{\text{th}}$ biomarker of a single cell in state $k$.   It is analogous to the spectroscopic observables of a protein molecule.  If $\vx_{\ell}=\big(x_{\ell}^{(1)},\cdots,x_{\ell}^{(K)}\big)$ with $\ell=1,\cdots,L$ are linearly independent vectors and $L>K$, then the $L$ biomarkers uniquely determine a cellular state.  If $L<K$, then the measurements contains uncertainty.  Of course there are always measurement errors.

The mean value of a biomarker among a large number of cells is analogous to the extensive macroscopic variable in thermodynamics.  Its measurement provides a {\em constraint} on the possible counting frequencies underlying the measurement.  Maximum entropy principle applies in this case to provide the most probable counting frequencies in the sample.  For more discussion see \cite{qian-cheng-qb}.

\section{Discussion}

Let us start by emulating J. A. Schellman \cite{schellman_bpc_97}:  ``Thermodynamics deals with the relations of empirical mean values, or sums, of measurable quantities and is a science which has no explicit dependence on the mechanistic level nature of a system.''  The mathematical procedure in Sec. \ref{sec:2} and Fig. \ref{fig01}, by only formally identifying the dependencies of $S_B(E)$ and $F(T)$ upon a $T$ and mechanical energy function $U(\vx)$ without further explicating any mechanical details, derives a statistical thermodynamics.  The mesoscopic, information entropy function introduced in (\ref{dFdT}a) with its two alternative expressions in (\ref{dFdT}b,c), however, requires an explicitly given $U(\vx)$, no matter how crude it is.  The entire art of protein science is an iterative duet between laboratory measurements, mostly in terms of concentrations (counting!) of different conformational states and coarse-grained energy function building, in terms of statistical weights or equilibrium constants, that includes conceptualizing the very state space $\mathcal{S}$ \cite{DBaker,qian_jpc_92}.

Gibbs' theory, with his statistical free energy based on $U(\vx)$ and a $T$, gives rise to  ``modern'' classical thermodynamic energetics in terms of a pair of dual variational principles and even the very concept of irreversible thermodynamic potential as a mathematical limit. The procedure can be formalized as $k_B\to 0$.  All the equilibrium thermodynamic relations taught in college classrooms in terms of multivariate calculus are consequences of the variational principles and their duality.  In a complete parallel, modern probability with its large deviations theory derives an Eulerian degree $1$ homogeneous entropy function for any random variable with probability distribution and its cumulant generating function, as in Eq. \ref{BoEn} \cite{qian_entropy_24}.  The neo-Gibbsian theory has been extended beyond i.i.d. statistical data \cite{cqy_cpb}, even leading to classical mechanics itself \cite{MQW-II}.  The duality symmetry makes the theory authentic with inner truth and sincerity \cite{puett-book}; it is a piece of true art.

\section{Acknowledgement}
We thank Professor Xiaofeng Jin (Fudan Univ.) for continuous discussions.  The second author dedicates this work to the memory of Jie Liang (1964-2024), a dear friend and a coauthor \cite{liangjie}, and to the memory of Professor John Schellman (1924-2014), the mentor who had taught him everything about Gibbs' thermodynamic energetics and applications to proteins.

%%%%%%%%%%%%%%%%%% BIBLIOGRAPHY %%%%%%%%%%%%%%%%%%
\input{reference.bbl}
\bibliography{reference}

\appendix
\section{Boltzmann's micro-canonical approach}
\label{app_A}

Identifying the $k_B\to 0$ limit with classical thermodynamics is inspired by the mathematics known as the  G\"{a}rtner-Ellis theorem \cite{touchette_phys_rep}.
Sec. \ref{sec:2} follows the isothermal canonical ensemble theory with a given temperature $T$.  It is possible to directly analyze Boltzmann's entropy in Eq. (\ref{BoEn}a), known as the micro-canonical approach.   In this case, one identifies the state space $\mathcal{S}_N$ as an $N$-dimensional subset in $\mathbb{R}^N$ where the ``degrees of freedom'' $N\to\infty$ as the macroscopic thermodynamic limit.  Therefore one has for large $N$
\begin{eqnarray}
  S_B(E,V,N) &=& k_B\log\left(\frac{1}{\rd E}\int_{
     \{E<U(\vx)\le E+\rd E\}\bigcap\mathcal{S}_N }
       \rd\vx \right)
\nonumber\\[6pt]
    &\simeq& Nk_B\log\Xi\left(\frac{E}{N},
     \frac{V}{N}\right),
\label{a1}
\end{eqnarray}
in which $V=\|\mathcal{S}_N\|$ is the phase volume and
\begin{equation}
\Xi(\eta,v) = \lim_{N\to\infty}
    \frac{1}{N}\log\left(\frac{1}{\rd E}\int_{
     \{N\eta<U(\vx)\le N\eta+\rd E\}\bigcap\mathcal{S}_N }
       \rd\vx \right),
\label{Boltz_LDP}
\end{equation}
where $\|\mathcal{S}_N\|=V=Nv$.  Formally, the macroscopic thermodynamic limit is $N\to\infty$, $k_B\to 0$, with fixed $Nk_B$.  This guarantees the existence of the macroscopic $S_B$ in (\ref{a1}).  The same assumption was implicitly made in (\ref{macro_limit}) when taking $k_B\to 0$ while holding $S_B(E)$ constant.

In modern probability theory, Eq. (\ref{Boltz_LDP}) is known as {\em large deviations principle} \cite{dembo-book}. Macroscopic entropy in (\ref{a1}) is an Eulerian degree $1$ homogeneous function of $E$, $V$, and $N$ \cite{callen-book}.  Then starting from a homogeneous $S(E,V,N)$, Helmholtz and Gibbs free energies follow LF transforms:
\begin{eqnarray}
    &\displaystyle
    F(T,V,N) = \min_{E}\Big\{E-TS(E,V,N)\Big\},
\label{TP-1}\\
    &\displaystyle
    G(T,p,N) = \min_{E,V}
     \Big\{E+pV-TS(E,V,N)\Big\},
\end{eqnarray}
and finally,
\begin{widetext}
\begin{equation}
\label{TP-3}
    \min_{E,V,N}
     \Big\{E+pV-\mu N-TS(E,V,N)\Big\}
     =\left\{\begin{array}{ccc}
       0  &&  \mu-\frac{\partial G}{\partial N}(T,p) \le 0  \\[5pt]
       -\infty && \text{elsewhere} \end{array}\right.
\end{equation}
\end{widetext}
where $G(T,p,N)$ is necessarily a linear function of $N$: $\frac{G}{N}=\frac{\partial G}{\partial N}$ is a function of $T$ and $p$ only \cite{dongwei_nature_23}.  Functions $\{\cdots\}$ in (\ref{TP-1})-(\ref{TP-3}) are three macroscopic irreversible thermodynamic potential functions for systems, respectively, under fixed $(T,V,N)$ with varying $E$, fixed $(T,p,N)$ with varying $E,V$, and fixed $(T,p,\mu)$ with varying $E,V,N$. For mesoscopic systems, the same functions provide descriptions for equilibrium fluctuations.

For a non-macroscopic system, $S(E,V,N)$ is not homogeneous.  Then
(\ref{TP-3}) becomes
\begin{eqnarray}
       \mathscr{E}(T,p,\mu) &=& \min_{E,V,N}
     \Big\{E+pV-\mu N-TS(E,V,N)\Big\}
\label{hgde}\\
    &=& \min_{N}
     \Big\{G(T,p,N)-\mu N\Big\}.
\nonumber
\end{eqnarray}
This is the subdivision potential in nanothermodynamics \cite{hill_nanolett}, from which Hill-Gibbs-Duhem equation follows, $\rd(\mathscr{E}/T) =E\rd(1/T)+V\rd(p/T)-N\rd(\mu/T)$ \cite{lu-qian-22}.  The function in $\{\cdots\}$ in (\ref{hgde}) represents the fluctuations among a large number of replicas of a small thermodynamic system under given $T,p,\mu$.  $\mathscr{E}(T,p,\mu)$ degenerates into (\ref{TP-3}) when $S(E,V,N)$ becomes Eulerian homogeneous.  The best example to which this theory applies is protein molecules in an aqueous solution, as complex systems \cite{hill_jpc_55}. In protein biophysics $\mathscr{E}$ was independently formulated by J. Wyman as binding potential in terms of a polynomial of ligand activity $a=e^{\mu/k_BT}$ \cite{wyman,wyman_gill} and carefully studied by Schellman and others \cite{schellman_75,Schellman_78} via semigrand partition function.

\end{document}

%% file: macros.tex
\def\rd{{\rm d}}

\def\vn{{\bf n}}

\def\vu{{\bf u}}

\def\vx{{\bf x}}